# WORDS FOR NOBEL PRIZES

JOSÉ M. MORÁN-MIRABAL

jmmoranm@ipicyt.edu.mx

*Departamento de Biología Molecular del IPICYT, Apdo. Postal 3-74, Tangamanga, San Luis Potosí, MÉXICO*

HARET C. ROSU

hcr@ipicyt.edu.mx

*Departamento de Matemáticas Aplicadas del IPICYT, Apdo. Postal 3-74, Tangamanga, San Luis Potosí, MÉXICO*

Date: March 20, 2002

We present the statistics of the significant nouns and adjectives of social impact figuring in the nominations of the Nobel prizes in Physics and Chemistry over the period of the awards from 1901 to 2001.

Interesting statistics on the ensemble of Nobel laureates can be found in the literature. [1] For the case of Physics, Zhang and Fuller, [2] provided a recent one that we strongly recommend to the reader. One can find therein percentage histograms and pie charts of Nobel winners in Physics according to their nationality, countries of final degree, age groups, religion, subfields of Physics, and so on. In the following, we review how the original aim of Alfred Nobel that his prize should be awarded to 'the person who has made the most important *discovery* or *invention* in the domain of X", where X stands for Physics, Chemistry and Medicine has been respected through the years. We thus perform a statistical study over the words employed in the awarding phrases for Physics (165 laureates) and Chemistry (137 laureates). We set aside Medicine (175 laureates), because for this science the word discovery (discoveries) is missing in only 4 awards after 1918. The results are presented in two tables and five histograms for 21 key nouns and 5 key adjectives of high social meaning as distributed over two epochs, before and after 1950. *Discovery (discoveries)* is indeed the most important word with 46(17) and 17(5) appearances in Physics and Chemistry, respectively. However, *invention* with 11 and 5 appearances, respectively, is surpassed by other words such as *contribution*, *development*, *method*, and *work*, the latter being the most frequent word in Chemistry with 22 appearances, equal to the sum of discovery and discoveries.

In the case of adjectives, the most frequent is *fundamental*, occurring essentially after 1950, with 10 and 7 appearances in Physics and Chemistry, respectively.

**Tables: Nouns and adjectives in alphabetical order with their total distribution and the division into before and after 1950 epochs.**

| Nouns | Physics | Until 1950 | After 1950 | Chemistry | Until 1950 | After 1950 |
|---|---|---|---|---|---|---|
| achievement | 1 | 0 | 1 | 3 | 0 | 3 |
| advancement | 1 | 1 | 0 | 3 | 3 | 0 |
| contribution | 12 | 2 | 10 | 16 | 4 | 12 |
| demonstration | 3 | 1 | 2 | 1 | 1 | 0 |
| determination | 1 | 0 | 1 | 7 | 2 | 5 |
| development | 19 | 5 | 14 | 16 | 2 | 14 |
| discoveries | 17 | 5 | 12 | 5 | 2 | 3 |
| discovery | 46 | 25 | 21 | 17 | 10 | 7 |
| elucidation | 0 | 0 | 0 | 3 | 0 | 3 |
| invention | 11 | 3 | 8 | 5 | 3 | 2 |
| method | 16 | 7 | 9 | 15 | 5 | 10 |
| preparation | 0 | 0 | 0 | 1 | 1 | 0 |
| recognition | 8 | 8 | 0 | 14 | 14 | 0 |
| research | 7 | 3 | 4 | 11 | 7 | 4 |
| service | 9 | 9 | 0 | 10 | 10 | 0 |
| studies | 6 | 1 | 5 | 6 | 0 | 6 |
| study | 4 | 1 | 3 | 2 | 2 | 0 |
| synthesis | 1 | 0 | 1 | 11 | 4 | 7 |
| theories | 1 | 0 | 1 | 1 | 0 | 1 |
| theory | 8 | 1 | 7 | 6 | 1 | 5 |
| work | 15 | 7 | 8 | 22 | 11 | 11 |

| Adjectives | Physics | Until 1950 | After 1950 | Chemistry | Until 1950 | After 1950 |
|---|---|---|---|---|---|---|
| basic | 2 | 0 | 2 | 0 | 0 | 0 |
| experimental | 5 | 2 | 3 | 1 | 0 | 1 |
| fundamental | 10 | 0 | 10 | 7 | 2 | 5 |
| pioneering | 7 | 0 | 7 | 1 | 0 | 1 |
| theoretical | 7 | 3 | 4 | 1 | 0 | 1 |

| # laureates | 165 | | | 137 | | |

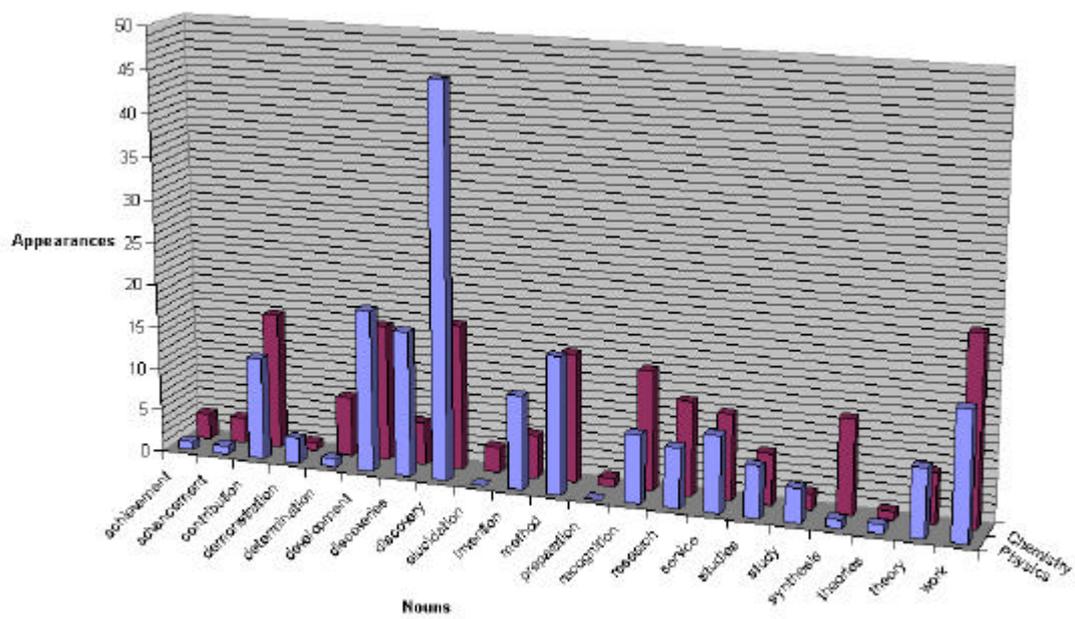

**Figure 1. Histograms of the total distribution of nouns for Physics and Chemistry, respectively.**

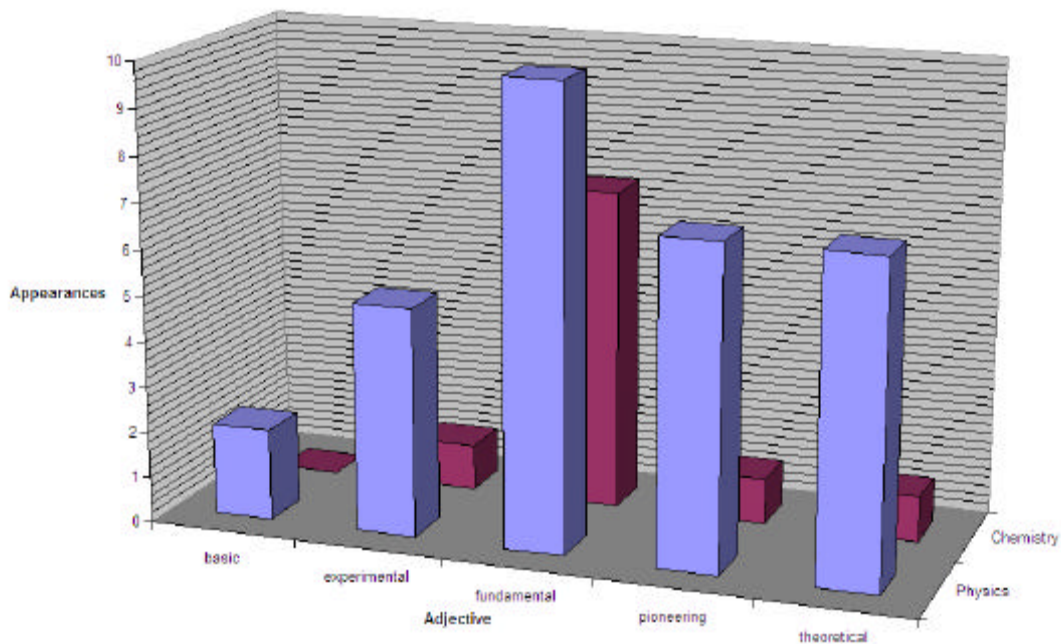

**Figure 2. Histograms of the total distribution of adjectives for Physics and Chemistry, respectively.**

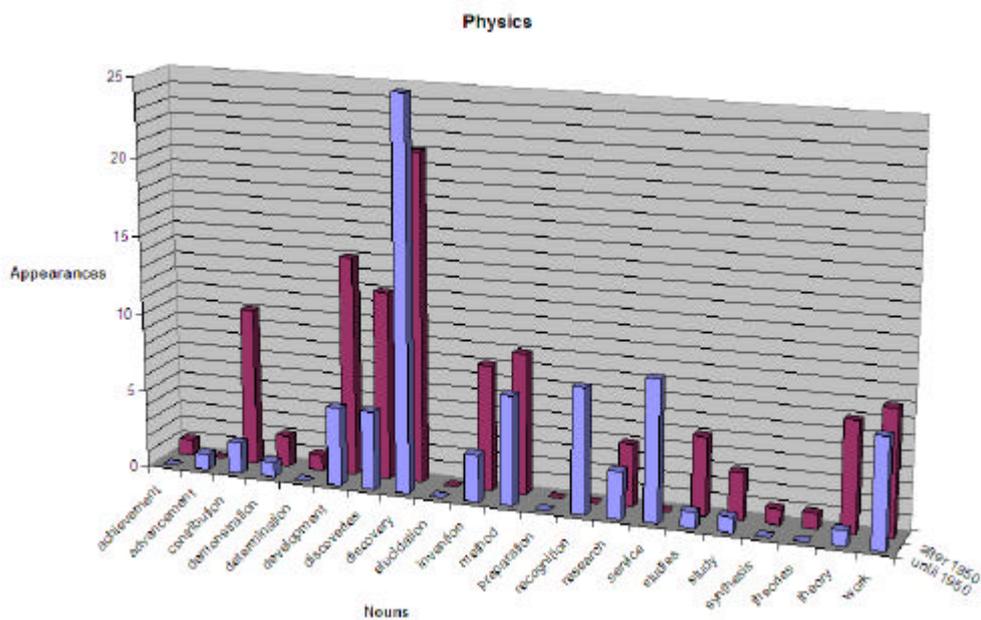

**Figure 3. Histograms of the half-epoch distributions of nouns for Physics.**

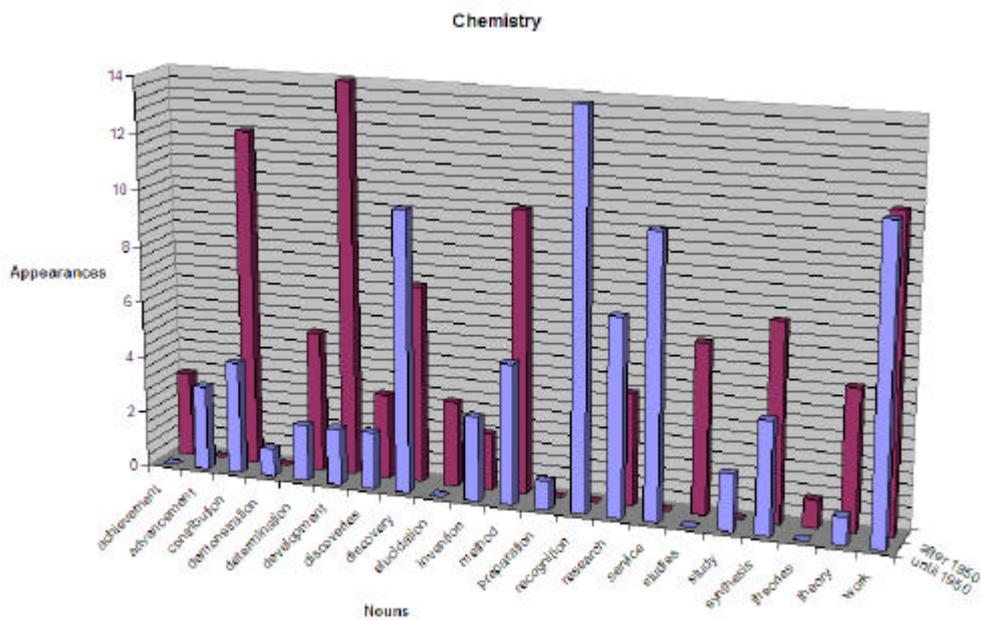

**Figure 4. Histograms of the half-epoch distributions of nouns for Chemistry.**

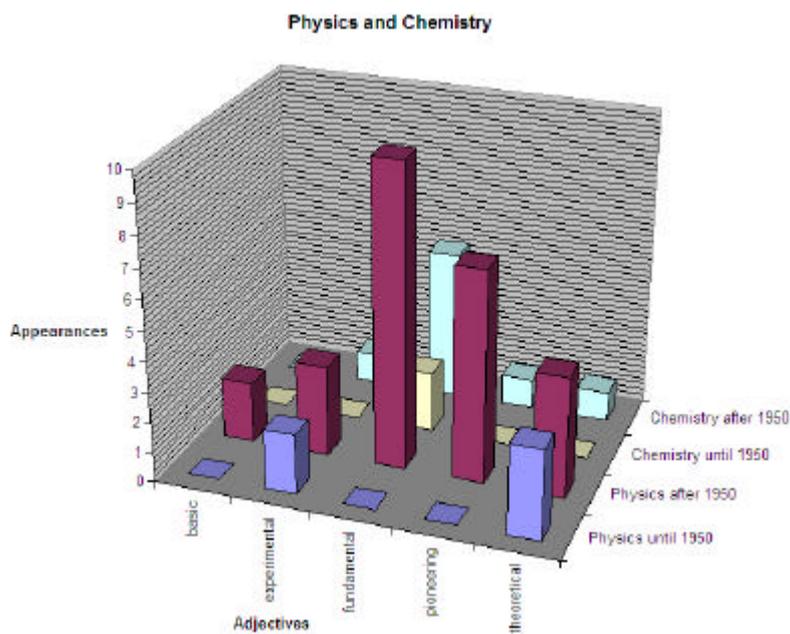

**Figure 5. Histograms of the half-epoch distributions of adjectives for both Physics and Chemistry.**